# Stacked Ensemble Learning for Abdominal Aortic Aneurysm Segmentation in CT Angiography

Joshua Fry[1[0009-0005-0002-6037]], Sajjad Arzemanzadeh[1[0000-0001-7381-1777]], Saeideh Sekhavat[1[0009-0006-9280-2781]], Mostafa Jamshidian[1[0000-0002-5166-171X]], Adam Wittek[1[0000-0001-9780-8361]], Michael Bertolacci[2[0000-0003-0317-5941]], Elke R. Gizewski[3[0000-0001-6859-8377]], Eva Gassner[3[0000-0002-8309-2638]], Alexander Loizides[3[0000-0001-5179-2309]], Maximillian Lutz[3[0009-0009-8141-8006]], Florian K. Enzmann[4[0000-0002-7200-4145]], Karol Miller[1[0000-0002-6577-2082]]

[1] Intelligent Systems for Medicine Laboratory, The University of Western Australia, Perth, Western Australia, Australia
[2] School of Physics, Mathematics and Computing, The University of Western Australia, Perth, Western Australia, Australia
[3] Department of Radiology, Medical University of Innsbruck, Innsbruck, Austria
[4] Department of Vascular Surgery, Medical University of Innsbruck, Innsbruck, Austria
mostafa.jamshidian@uwa.edu.au

**Abstract.** Abdominal aortic aneurysm (AAA) rupture risk assessment increasingly relies on patient-specific biomechanical computations, which require accurate three-dimensional aneurysm geometry from computed tomography angiography (CTA). Manual and semi-automated segmentation remain time-consuming and observer-dependent, limiting their use in large-scale clinical workflows. In this study, we developed a stacked ensemble framework for automated AAA segmentation from CTA images. We used 40 anonymised contrast-enhanced CTA scans from AAA patients and generated reference segmentations using the nnInteractive extension in 3D Slicer. We partitioned the dataset into 32 training cases and 8 held-out test cases. Three nnUNetv2 configurations, Default, DA5, and ResEncL, were trained as base learners, and their voxel-wise probability outputs were combined using an L2-regularised logistic regression meta-model trained from out-of-sample cross-validation predictions. We evaluated segmentation performance using Dice Coefficient and Separation Distance, a mean boundary-to-boundary distance measure introduced in this study to quantify average surface agreement. On the held-out test set, the ensemble achieved the highest mean Dice Coefficient of 0.9752 and the lowest mean Separation Distance of 0.4598 mm, indicating improved volumetric overlap and average boundary agreement compared with the individual base learners. Overall, stacked ensemble learning provided small but meaningful improvements in AAA segmentation, particularly for boundary accuracy relevant to downstream patient-specific biomechanical computations.

## 1 Introduction

Abdominal aortic aneurysm (AAA) is a typically asymptomatic vascular condition characterised by dilation of the abdominal aorta to a diameter exceeding 3.0 cm, approximately two standard deviations above the average aortic diameter of 1.8 cm [1, 2]. AAA develops due to structural weakening of the aortic wall and ruptures when blood pressure-induced wall stress exceeds wall strength. Although AAA may remain clinically silent, rupture is life-threatening, with mortality rates reaching up to 80% without immediate intervention [3]. Current clinical guidelines recommend surgical intervention when the maximum aneurysm diameter exceeds 5.5 cm in men and 5.0 cm in women, or when the annual growth rate exceeds 1 cm [1, 4]. However, aneurysm diameter alone is an unreliable predictor of rupture risk at the individual level. Many aneurysms exceeding clinical thresholds remain stable throughout a patient's lifetime [1, 4, 5], while smaller aneurysms may still rupture [6]. Autopsy studies report that approximately 60% of AAAs larger than 5 cm do not rupture, whereas nearly 13% of ruptured AAAs measure 5 cm or less [7].

Towards patient-specific AAA assessment, biomedical computational studies have investigated AAA wall stress [8-14], strain [15-17], and more recently, structural integrity [18, 19]. These computations typically require accurate three-dimensional aneurysm geometry reconstructed from CTA images [20-23], making reliable image segmentation a critical prerequisite for clinically meaningful biomechanical analysis [24]. Manual segmentation is time-consuming and labour-intensive, typically requiring 2–4 hours per patient. Semi-automated approaches reduce processing time to approximately 45 minutes but still require expert input. Both approaches are also subject to intra- and inter-observer variability, limiting scalability and reproducibility. Hence, fast, accurate, and fully automated segmentation methods are required to support large-scale clinical deployment of patient-specific biomechanical analysis.

Deep learning methods, particularly convolutional neural networks (CNNs), have become the dominant approach for medical image segmentation, achieving state-of-the-art performance across a wide range of applications [25-27]. The introduction of U-Net in 2015 marked a major advancement by providing an encoder–decoder architecture with skip connections that enabled accurate segmentation on the relatively small datasets typical of medical imaging [27]. Building on this foundation, nnUNet was introduced as a self-configuring framework that automatically adapts preprocessing, architecture, training, and postprocessing pipelines to a given dataset [25]. Combined with extensive data augmentation strategies, nnUNet has consistently achieved state-of-the-art performance across diverse medical segmentation tasks [28, 29]. In AAA segmentation, nnUNet has demonstrated strong performance, including success in recent challenges such as AortaSeg24 [30]. The availability of multiple nnUNet configurations with differing architectural and training characteristics also provides opportunities to exploit model diversity for improved segmentation performance.

Ensemble learning is a widely used machine learning technique that combines multiple models (base learners) to improve prediction accuracy and generalisation performance compared to any single model [31-33]. Among ensemble strategies, stack-ing is particularly well suited to segmentation tasks, where independently trained base learners provide voxel-wise probability maps that are combined by a meta-model. This enables the meta-model to exploit complementary strengths and fine-grained differences between base learners. In nnUNet, probabilistic voxel-level outputs are directly accessible, enabling their use within stacking-based ensembles [25, 30, 34]. Since different nnUNet configurations may capture complementary features through variations in architecture, augmentation, and training strategy, stacking offers strong potential for improving AAA segmentation performance. Nevertheless, the effectiveness of stacking can be limited by correlations between base learner outputs, motivating the use of regularised meta-models to improve generalisation.

In this study, we construct a stacked ensemble segmentation framework that combines three nnUNetv2 configurations using a regularised logistic regression meta-model. The paper is organized as follows. In Section 2, we present the AAA patient image dataset, training data preparation, base learner configurations, and the proposed stacking framework. In Section 3, we present the segmentation performance results and comparative analysis, followed by the discussion in Section 4 and conclusions in Section 5.

# 2 Materials and Methods

## 2.1 Dataset and Ground Truth Segmentation

We used a dataset of 40 anonymised contrast-enhanced CTA image datasets from 40 AAA patients. Patients were recruited at the Medical University of Innsbruck (Innsbruck, Austria), and informed consent was obtained prior to participation. The study was conducted in accordance with the Declaration of Helsinki, and the protocol was approved by the Ethics Committee of Medical University of Innsbruck (approval code 1271/2023).

We generated ground truth segmentations of the AAA structures using the nnInteractive extension [35] within the 3D Slicer platform [36]. nnInteractive is a 3D promptable AI-based segmentation tool that converts sparse user inputs, such as points and scribbles, into high-quality volumetric segmentations. Unlike traditional slice-by-slice approaches, it directly generates three-dimensional label maps from sparse 2D prompts, improving annotation efficiency and consistency while still incorporating expert guidance. Fig. 1 shows a representative CTA image, the corresponding AAA segmentation, and the resulting three-dimensional visualization of the segmented aneurysm structure.

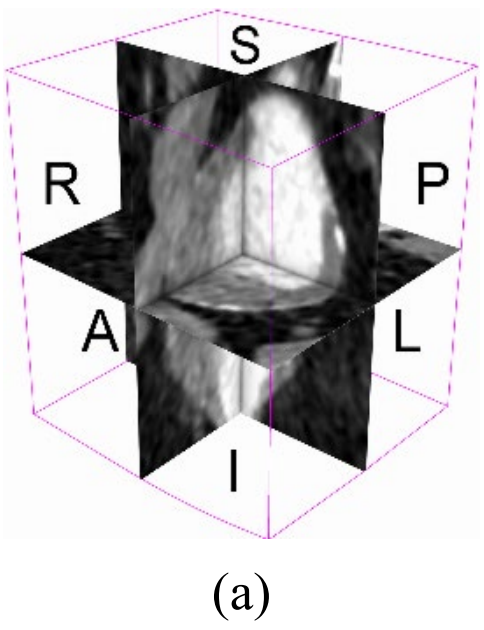

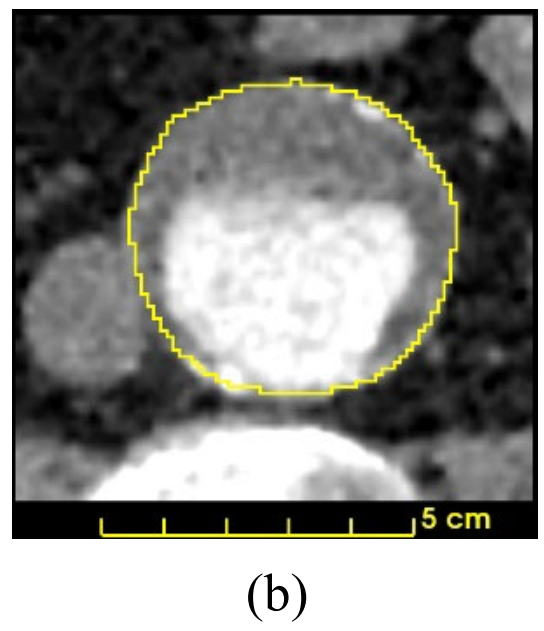

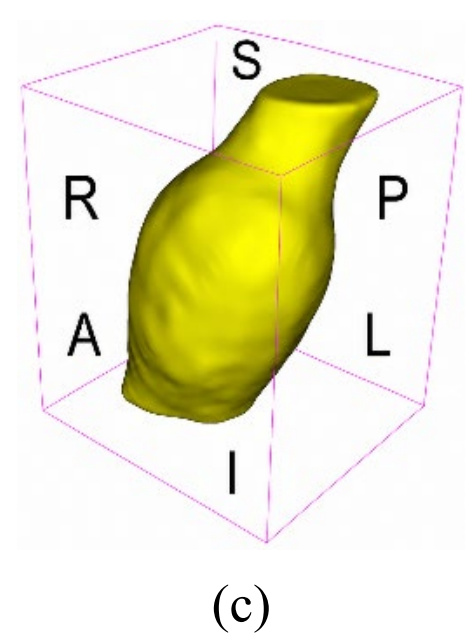

(a) (b) (c)

**Fig. 1.** Example of abdominal aortic aneurysm (AAA) image acquisition, segmentation, and three-dimensional visualization. (a) Contrast-enhanced CTA image displayed in multiplanar view for a representative patient. (b) Typical cross-sectional slice on a representative RA plane showing the AAA segmentation generated using the nnInteractive extension within the 3D Slicer platform. (c) Three-dimensional visualization of the segmented AAA structure reconstructed from the volumetric segmentation.

### 2.2 Dataset Partitioning and Cross-Validation

We divided the dataset into training and testing subsets to enable unbiased model evaluation. We randomly selected 8 images as a held-out test set and used the remaining 32 images for model development. Fig. 2 shows three-dimensional visualizations of the 32 AAA training segmentations used for model development.

To maximize the use of the available training data, we employed four-fold cross-validation, with each fold consisting of 24 training images and 8 validation images. For each fold, we trained the base learners on the training subset and generated out-of-sample predictions on the corresponding validation subset. This process produced out-of-sample predictions for all 32 training images from each base learner. We then used these predictions to train the ensemble meta-model, ensuring that the meta-model learned only from predictions generated on unseen data. After cross-validation, we retrained the base learners on the full set of 32 training images and performed final evaluation on the 8 held-out test images.

### 2.3 Framework overview

The proposed framework employs a stacked ensemble segmentation pipeline, illustrated in Fig. 3. We independently trained three nnUNetv2 base learners using different network configurations: Default nnUNet, nnUNet DA5, and nnUNet ResEncL. Each base learner generated voxel-wise probability maps for AAA segmentation. We then used these probability outputs as input features to a regularised logistic regression meta-model, which learned to combine the base learner predictions and produce the final segmentation output.

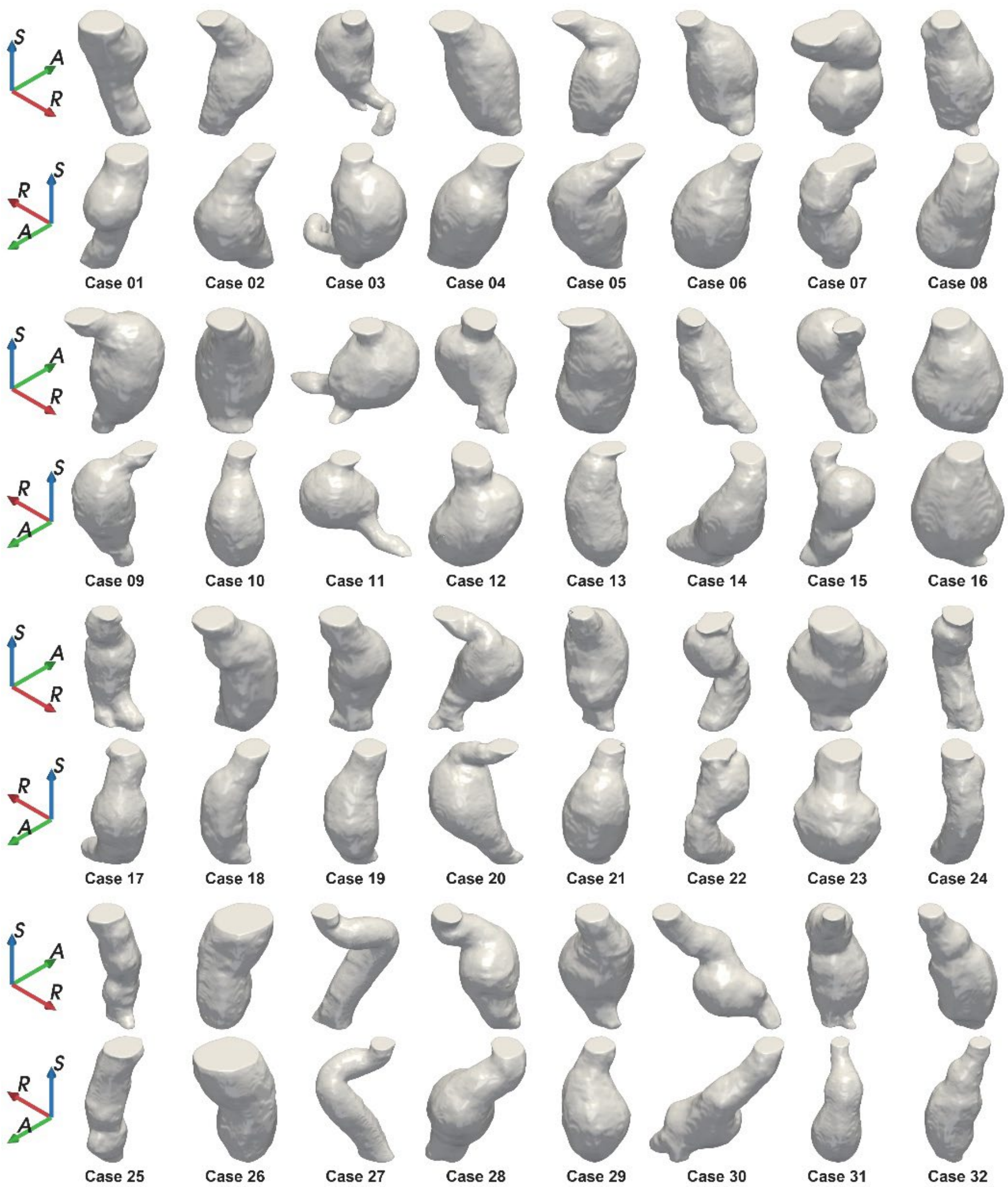

**Fig. 2.** Visualizations of the 32 abdominal aortic aneurysm (AAA) ground truth segmentations used for model training and cross-validation. Each case is shown in two different views to illustrate the anatomical variability in aneurysm morphology across the dataset.

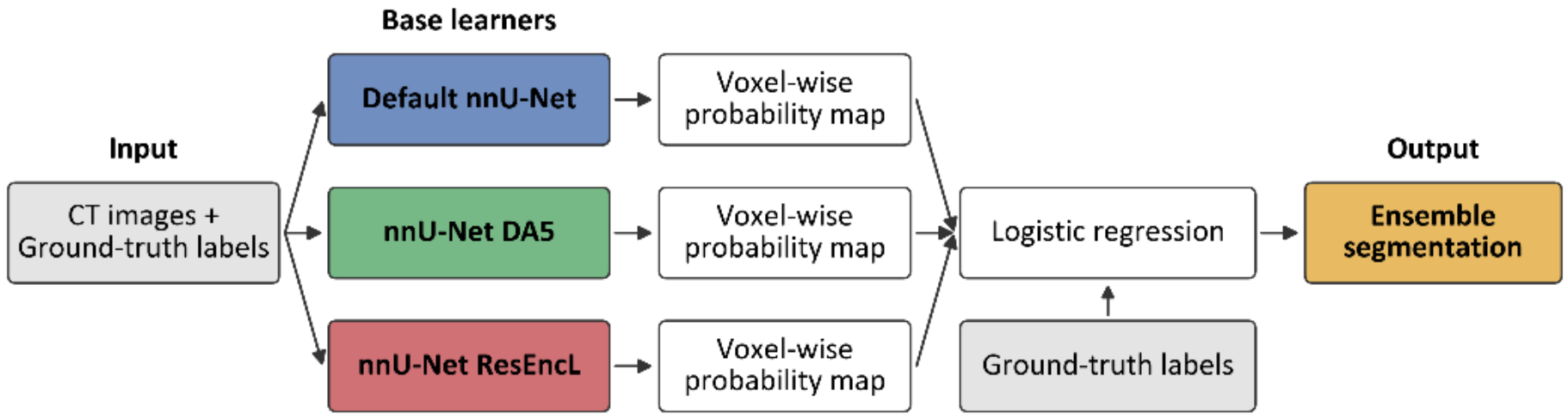

**Fig. 3.** Overview of the proposed stacked ensemble segmentation framework. Three nnUNetv2 base learners (Default, DA5, and ResEncL) are trained independently to generate voxel-wise probability maps, which are subsequently combined using a regularised logistic regression meta-model to produce the final segmentation. Gray: ground-truth, blue = Default nnUNetv2, green = nnUNetv2 DA5, red = nnUNetv2 ResEncL, and orange = ensemble model.

### 2.4 Base learners

The ensemble framework used three nnUNetv2 base learner configurations: Default, DA5, and ResEncL. These configurations differ in architecture and training strategy, providing diversity in the learned feature representations. The DA5 configuration applies a more extensive data augmentation pipeline than the default nnUNet, whereas the ResEncL configuration employs a larger encoder to capture more complex features. We trained all models on an NVIDIA GeForce RTX 4090 GPU. The implementation details and architectural specifications for each configuration follow the nnUNet framework described in [37].

### 2.5 Stacked Meta-Model

**Logistic Regression Formulation.** We employed stacking as the ensemble strategy, where independently trained base learners generated outputs that served as input features to a meta-model [31]. Because segmentation can be formulated as a voxel-wise classification problem, we selected logistic regression [38] as the meta-model.

Let $\boldsymbol{x}_i$ denote the feature vector for voxel $i$, consisting of the probability outputs from the base learners, and let $y_i \in \{0,1\}$ denote the corresponding ground truth class label. The logistic regression model estimates the probability of class membership as

$$P(y_i = 1 \mid x_i) = \frac{1}{1 + \exp(-\beta^T x_i)} \quad (1)$$

where β represents the model coefficients.

**Regularization.** A major challenge in stacking arises from the high correlation between base learner outputs. To reduce overfitting, we applied L2 regularization to the logistic regression model. The total loss function is defined as

$$L(\boldsymbol{\beta}) = L_{log}(\boldsymbol{\beta}) + \lambda \, ||\boldsymbol{\beta}||_2^2 \quad (2)$$

where $L_{log}$ is the logistic loss function (log-loss) and $\lambda$ is the regularization parameter. The regularization term penalises large coefficient values and improves generalisation performance.

**Hyperparameter Tuning.** We selected the regularization parameter $\lambda$ using leave-one-out cross-validation (LOOCV) [38]. For each candidate value of $\lambda$, we trained the model on all but one sample and evaluated it on the held-out sample. We then selected the value of $\lambda$ that minimized the average log-loss across all folds. Finally, we trained the meta-model on all available training data using the optimal value of $\lambda$ and used it to generate predictions on the test set.

**Evaluation Metrics.** We evaluated segmentation quality using the Dice Coefficient (DC) [39] and Separation Distance (SD). Let $A$ denote the predicted segmentation and $B$ denote the ground truth segmentation.

The Dice Coefficient measures the overlap between A and B:

$$\mathrm{DC}(A, B) = \frac{2|A \cap B|}{|A| + |B|} \tag{3}$$

To quantify boundary proximity, we used Separation Distance, defined as the mean shortest distance from the ground truth boundary points to the predicted boundary:

$$\mathrm{SD}(A, B) = \frac{1}{n}\sum_{i=1}^{n} \delta_i(A, B) \tag{4}$$

where $\delta_i(A, B)$ denotes the shortest distance from the $i$-th ground truth boundary point in $B$ to the boundary of the predicted segmentation $A$, and $n$ is the total number of ground truth boundary points. This one-directional formulation measures how closely the predicted boundary follows the reference AAA surface. Separation Distance focuses only on distances from the ground truth boundary to the predicted boundary. Therefore, it specifically evaluates the proximity of the prediction to the reference anatomy and is less influenced by small additional predicted boundary regions. Lower SD values indicate closer agreement between the predicted and ground truth boundaries.

### 2.6 Experiments

We conducted experiments to evaluate whether the proposed stacked ensemble framework improves segmentation performance compared with the individual nnUNetv2 configurations described in Section 2. In particular, we investigated whether combining complementary nnUNetv2 models through stacking improves segmentation overlap and boundary accuracy, as measured by the Dice Coefficient and Separation Distance.

We evaluated the proposed ensemble framework against the three base learner configurations: Default nnUNetv2, nnUNetv2 DA5, and nnUNetv2 ResEncL. These models provide strong baselines, as nnUNet is widely recognized as a state-of-the-art segmentation framework [25, 28].

We evaluated model performance on the held-out test set of 8 CTA images described in Section 2. For each test image, we generated segmentation outputs for all base learners and the ensemble model. We quantified performance using the Dice Coefficient and Separation Distance, and computed aggregate performance by averaging the metric values across all test images.

## 3 Results

Fig. 4 presents qualitative comparisons between the ground truth segmentations and the segmentations generated by the three nnUNetv2 base learners and the proposed ensemble model for the 8 held-out test cases. Visual inspection shows that all models successfully capture the overall aneurysm morphology and produce segmentations that closely match the ground truth surface. The ensemble model achieves segmentation quality comparable to the strongest individual base learners, with slight improvements in boundary consistency in some anatomically complex regions.

We first evaluated the three nnUNetv2 base learners using out-of-fold predictions across the 32 training images, as summarized in Table 1. The reported mean values represent averages across these 32 training images. The nnUNetv2 DA5 configuration achieved the highest mean Dice Coefficient (0.9704) and the lowest mean Separation Distance (0.5449 mm) among the base learners, indicating the best overlap and boundary proximity to the ground truth. The Default nnUNetv2 achieved comparable Dice performance (0.9675), while the ResEncL configuration showed slightly lower Dice performance (0.9634). Overall, these results show measurable variability across the base learner configurations, supporting their combination through ensembling.

We then evaluated the ensemble model on the held-out test set of 8 images, as summarized in Table 2. The reported mean values represent averages across these 8 test images. The ensemble achieved the highest mean Dice Coefficient (0.9752) and the lowest mean Separation Distance (0.4598 mm), outperforming all individual base learners in both metrics. This indicates improved volumetric overlap and closer agreement with the ground truth boundary. These quantitative findings are consistent with the qualitative observations in Fig. 4, where the ensemble showed segmentation quality comparable to the strongest individual models, with slightly improved boundary consistency in some cases.

Overall, the ensemble demonstrated modest but consistent improvements in aggregate segmentation performance, particularly in Dice Coefficient and Separation Distance, supporting the effectiveness of stacked ensemble learning for automated AAA segmentation from CTA images.

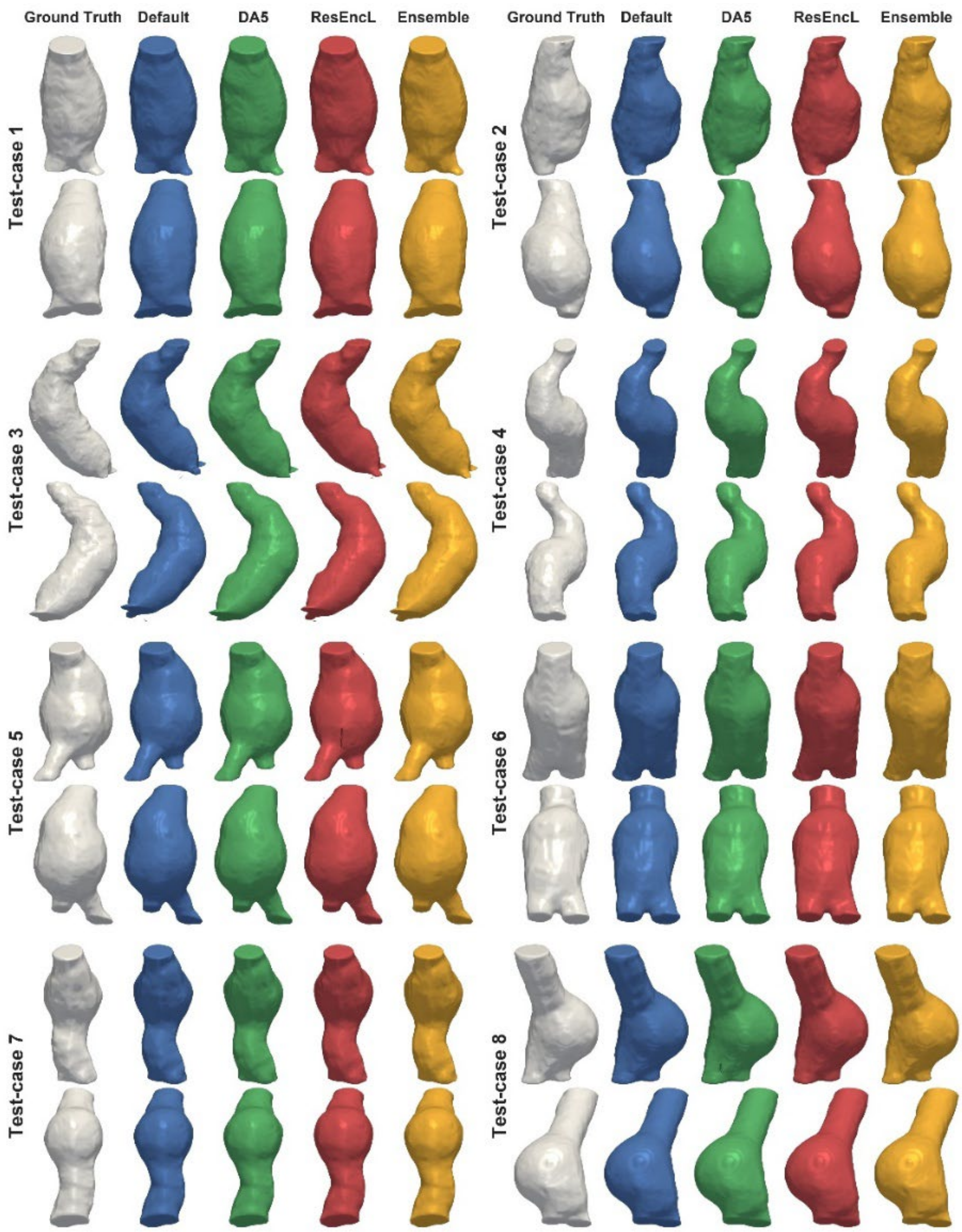

**Fig. 4.** Qualitative comparison of segmentation results for the 8 held-out test cases. For each case, the ground truth segmentation is compared with the segmentations generated by the Default nnUNetv2, nnUNetv2 DA5, nnUNetv2 ResEncL, and ensemble models. The ensemble model demonstrates segmentation quality comparable to the strongest individual base learners, with slight improvements in boundary consistency observed in some anatomically complex regions. Colours are consistent with Fig. 3: Gray: ground-truth, blue = Default nnUNetv2, green = nnUNetv2 DA5, red = nnUNetv2 ResEncL, and orange = ensemble model.

**Table 1.** Performance of nnUNetv2 base learners on the training data using out-of-fold evaluation. Values represent the mean Dice Coefficient and mean Separation Distance (mm) across the 32 training images.

| Model | Mean Dice Coefficient | Mean Separation Distance (mm) |
|---|---|---|
| Default nnUNetv2 | 0.9675 | 0.5806 |
| nnUNetv2 DA5 | 0.9704 | 0.5449 |
| nnUNetv2 ResEncL | 0.9634 | 0.5850 |

**Table 2.** Segmentation performance on the held-out test set. Values represent the mean Dice Coefficient and mean Separation Distance (mm) across the 8 test images. The ensemble achieves improved aggregate performance, particularly in Dice Coefficient and Separation Distance.

| Model | Mean Dice Coefficient | Mean Separation Distance (mm) |
|---|---|---|
| Default nnUNetv2 | 0.9748 | 0.4653 |
| nnUNetv2 DA5 | 0.9744 | 0.4763 |
| nnUNetv2 ResEncL | 0.9743 | 0.4707 |
| Ensemble Model | 0.9752 | 0.4598 |

# 4 Discussion

This study showed that the proposed stacked ensemble model provided modest improvements over the individual nnUNetv2 base learners for abdominal aortic aneurysm segmentation. The ensemble achieved the highest mean Dice Coefficient of 0.9752 and the lowest mean Separation Distance of 0.4598 mm on the held-out test set. These results suggest that combining the voxel-wise probability outputs of multiple nnUNetv2 configurations can improve overall segmentation accuracy, particularly in terms of volumetric overlap and boundary agreement.

The improvement was relatively small, which is expected because all individual nnUNetv2 models already performed at a high level. When base learners are strong, the remaining segmentation error is limited, so large performance gains from ensembling are unlikely. Nevertheless, the ensemble was able to combine complementary predictions from the Default, DA5, and ResEncL configurations. Differences in architecture, augmentation strategy, and feature extraction capacity likely allowed each model to capture slightly different aspects of the aneurysm anatomy, especially near uncertain boundary regions.

The most notable improvement was observed in Separation Distance, indicating that the ensemble produced boundaries that were, on average, slightly closer to the ground truth segmentation. This is important for downstream applications such as finite element analysis, where accurate reconstruction of aneurysm geometry is essential for reliable biomechanical modelling. Even small improvements in boundary placement may be valuable when segmentation outputs are used for patient-specific simulations.

This study has several limitations. We used a relatively small dataset of 40 CTA scans, with only 8 cases reserved for testing. In addition, all images were collected from a single clinic and scanner, which may limit the generalisability of the findings to images acquired using different scanners and imaging protocols. However, the dataset produced strong segmentation results partly because the imaging field of view was focused on the AAA region. In more general clinical settings, CTA scans may include wider or more variable anatomical coverage beyond the aneurysm. Therefore, the proposed framework may require an initial localisation and cropping step before segmentation. By first identifying and cropping the region of interest around the aneurysm, the stacked ensemble model could then be applied to a focused AAA field of view like the images used in this study.

Overall, the results indicate that stacked ensemble learning can provide small but meaningful improvements for AAA segmentation from CTA images, especially when the segmentations are used for downstream patient-specific simulations. Future work should validate the method on larger and more diverse datasets and assess whether improved segmentation accuracy leads to more reliable computation-based patient-specific biomarkers.

## 5 Conclusion

In this study, we developed a stacked ensemble framework for abdominal aortic aneurysm segmentation from CTA images by combining three nnUNetv2 configurations using a logistic regression meta-model. The ensemble achieved the highest mean Dice Coefficient of 0.9752 and the lowest mean Separation Distance of 0.4598 mm on the held-out test set, indicating modest improvements in volumetric overlap and average boundary agreement compared with the individual base learners.

The results show that stacked ensemble learning can improve average AAA segmentation performance, even when the base learners already achieve high accuracy. The most relevant improvement was observed in Separation Distance, suggesting improved boundary placement, which is important for downstream patient-specific biomechanical computations.

The findings should be interpreted considering the relatively small, single-source dataset and the focused AAA field of view used in this study. Future work should validate the method on larger and more diverse datasets, incorporate an initial localisation and cropping step for scans with variable anatomical coverage, and assess whether improved segmentation accuracy leads to more reliable computation-based patient-specific biomarkers.